\documentclass[a4paper,fleqn]{cas-sc}

\usepackage[authoryear,longnamesfirst]{natbib}

\def\tsc#1{\csdef{#1}{\textsc{\lowercase{#1}}\xspace}}
\tsc{WGM}
\tsc{QE}

\usepackage{graphicx}
\usepackage{subfig}
\usepackage{amsmath}
\usepackage{mathtools}
\usepackage{multirow,booktabs}
\usepackage{ulem}
\everymath{\displaystyle}

\usepackage{soul}

\usepackage{color}

\renewcommand{\d}{{\rm d}}
\newcommand{\e}{{\rm e}}

\newcommand{\eq}{\begin{equation}}
\newcommand{\eqend}{\end{equation}}
\newcommand{\ray}{\begin{eqnarray}}
\newcommand{\rayend}{\end{eqnarray}}

\newcommand{\eqn}[1]{(\ref{#1})}

\newcommand{\PD}[2]{\frac{\partial#1}{\partial#2}}

\newcommand{\pv}{{\bf p}}
\newcommand{\Pv}{{\bf P}}
\newcommand{\rv}{{\bf r}}

\newcommand{\sv}{{\bf s}}

\newcommand{\Fv}{{\bf F}}
\newcommand{\Rv}{{\bf R}}

\newcommand{\Av}{{\bf A}}
\newcommand{\Bv}{{\bf B}}

\newcommand{\Ev}{{\bf E}}

\begin{document}
\let\WriteBookmarks\relax
\def\floatpagepagefraction{1}
\def\textpagefraction{.001}

\shorttitle{Approximate Approach to Coulomb Entanglement}    

\shortauthors{M.~Ballicchia, C.~Etl, M.~Nedjalkov,  D.K. Ferry,  H.~Kosina, J.~Weinbub }  

\title [mode = title]{  Approximate Wigner Approach to Coulomb Entanglement
}  

\tnotemark[1] 

\tnotetext[1]{} 

%

\author[1]{Mauro Ballicchia}[orcid=0000-0002-7391-2698,]


\ead{m.ballicchia@gmail.com}



\affiliation[1]{organization={Institute for Microelectronics, TU Wien},
            addressline={Gusshausstrasse 27-29}, 
            city={Vienna},
          postcode={1040}, 
            country={Austria}}

\author[1]{Clemens Etl}[orcid=0000-0001-7331-6338]

\cormark[1]
\cortext[1]{Corresponding author:  C.Etl }
\ead{etl@iue.tuwien.ac.at}



\ead[url]{}



\author[1]{Mihail Nedjalkov}[orcid=0000-0002-5705-251X]
\ead{mixi@iue.tuwien.ac.at}
\author[2]{David K. Ferry}[orcid=0000-0002-0942-8033]
 
\affiliation[2]{organization={School of Electrical, Computer, and Energy Engineering, Arizona State University},
            city={Tempe},
            postcode={85287}, 
            country={USA}}
\ead{ferry@asu.edu}
\author[1]{Hans Kosina}[orcid=0000-0003-1616-4942]
\ead{kosina@iue.tuwien.ac.at}

\author[3]{Josef Weinbub}[orcid=0000-0001-5969-1932]
\ead{josef.weinbub@silvaco.com}
\affiliation[3]{organization={Silvaco Europe Ltd.},
            city={St Ives},
            postcode={PE27 5JL}, 
            country={UK}
}


\begin{abstract}
The electric interaction between two nearby evolving electrons triggers the correlation between their waves and governs the operation of logical devices called Coulomb entanglers.
Of technological interest in the presence of magnetic fields are multi-spatial evolution scenarios beyond pure state descriptions. The two-electron density matrix becomes eight-dimensional even for two-dimensional spatial cases and is thus computationally prohibitive.
In this work, we present two new approximations of the two-electron Wigner equation that aim at computational feasibility: a BBGKY approach for reducing the number of variables and a field approximation of the Coulomb-Wigner operator. They exhibit different conceptual aspects that illustrate alternative viewpoints to entanglement: Only the evolution provided by the latter model satisfies the orthodox definition of entanglement. Our analysis, based on the Fredholm integral representation of the models, allows us to develop an intuitive picture and physical insight into the process.
\end{abstract}


\begin{highlights}
\item The computationally prohibitive two-electron problem of Coulomb entanglement is approached with Wigner (phase space) quantum mechanics.
\item Two approximate methods, which are well-established in the quantum transport theory, are now applied and analyzed with respect to their relevance to describe the process of entanglement.
\item Projections over  subspaces are used to obtain the BBGKY hierarchy of equations, which gives rise to a model violating the orthodox definition of entanglement.  However, we show that the evolution is non-linear and non-Markovian, demonstrating correlations between the reduced (single electron) states.
\item The field approximation of the Coulomb-Wigner operator gives rise to classical forces that govern the  trajectories' characteristics of the Liouville operator.  We show that the latter, being defined in the two-electron phase space, gives rise to entanglement.
\item
The paper focuses on the physical insights provided by the two derived approximate models on different aspects of the process of entanglement, while the mathematical derivations are given in the appendices.
\end{highlights}


\maketitle

\section{Introduction}

With the many astonishing advances in coherent single-electron systems (e.g., single-electron sources, electron interferometers, and flying charge qubit networks)~\cite{Edlbauer2022}, the obvious next step is to venture into entangled multi-electron systems~\cite{Hofer2017,Moskalets2020,Kotilahti2021,Edlbauer2022}.  
Historically, the focus of the research was on spin entanglement. Recently, however,  there has been growing interest in spatial entanglement via correlated electron waves, allowing for a particularly intuitive view of the phenomenon~\cite{Schroeder2017}. 
When electrons are in close proximity to each other, Coulomb interactions dominate and generate spatial entanglement between the electrons ("entanglement by interaction"), such as in Coulomb entanglers/waveguide couplings~\cite{Reichl2005,Abdullah2016,Olendski2018}, quantum dot systems~\cite{Pham2020}, electron-electron (e-e) collision studies~\cite{Gollisch2019}, quantum teleportation~\cite{Galler2021}, interferometers~\cite{Buscemi2012,Bellentani2019,Bordone2019}, and conditional phase shifters~\cite{Bellentani2020}. 
Modeling entangled multi-electron dynamics and spatial interference effects is, in general, a much-needed ability in the analysis of the involved processes. 
For instance, the antibunching of two interacting fermionic wave packets impinging on a quantum point contact has recently been studied within the context of Hong-Ou-Mandel (HOM) interferometers~\cite{Bellentani2019}. 
The transport region is a regular two-dimensional electron gas structure in the integer quantum Hall regime, and a perpendicular magnetic field is used to propagate the electrons. It was shown that modeling (i) the geometry in two-dimensional (rather than how it is usually done via simplified one-dimensional models) and (ii) the Coulomb interaction offered a crucial advantage for a correct physical analysis of the system. In addition, it was shown that Coulomb repulsion is much more dominant than the exchange energy. 

Entanglement has been called the most important aspect of quantum mechanics by E. Schr\"odinger, who coined the term almost a century ago \cite{Schroedinger35}.
The counterintuitive physical aspects of entanglement inspired discussions involving the EPR paradox and the phrase "spooky action at a distance" \cite {EPR}.
In essence, this is related to specific correlations between sub-systems of a composite quantum system:  
The state of the latter behaves as a whole and not as a result of the individual behavior of the components. Such a spatially extended composite state can be affected at one end and 'feel' this at the very distant other end.
A two-particle state defines two reduced single-particle states, obtained by tracing out one or the other set of single-particle degrees of freedom jointly describing the system. 
Of physical interest are scenarios where the two reduced states are localized away from each other so that affecting one of them also affects the other 'at a distance'.
This is illustrated by the e-e evolution in Coulomb entanglers. Two initially distant electrons in locations $1$ and $2$ evolve (and interact) together for a certain time and are then directed towards other distant locations $A$ and $B$. One can measure the state of an electron in $A$ and then know the corresponding state in $B$. This is because the interaction transforms the two initially single-electron 
states into a two-electron state: During the interaction, they lose their individuality and become tied with probability amplitudes.
The process is characterized by a loss of coherence, namely, the evolution of the single electrons is no longer coherent, giving rise to a reduction of their {states'} purity. 
Without interaction, we can associate a trajectory with each of the evolving electrons (e.g., using the Ehrenfest theorem) and thus trace it from the initial to the final place.
With interaction, we cannot determine which of the electrons, initially being in $1$ or $2$, falls in $A$ or $B$. 
This heuristic picture illustrates the physical viewpoint of entanglement.
The formal definition of entanglement is closely related to this picture. It suggests that a pure many-particle state is entangled if it is not separable into a tensor product of single-particle states. 
Such a definition, however, depends on the choice of the basis vectors.
The concept is further generalized with the help of the Schmidt decomposition, which provides a criterion for separability.
However, for mixed states, "similar tools are available only for low-dimensional systems"  \cite{Buch}.
"The search for a fully general definition of entanglement remains an active area of research” \cite{BJ}. 
In other words, despite its scientific and technological significance, "there is no consensus about how to define entanglement in a general way that transcends any specific realization"~\cite{Shaw}.
Furthermore, from a mathematical point of view, coherence and entanglement are closely linked:
"Quantum coherence and quantum entanglement are two sides of the same coin"~\cite{Z}. 
Another viewpoint links entanglement with coherence:
The recently developed resource theory of coherence is quantified by following the ideas of the quantification theory of entanglement~\cite{Baum}. 
This means that the two concepts describing different physical notions have a common mathematical foundation, which is based on the superposition property.
Furthermore, the two concepts are quantitatively equivalent, i.e., any nonzero amount of coherence in a system can be converted into an equal amount of entanglement between that system and another initially incoherent one~\cite{Strelt}. 
This means that quantum coherence can be measured through entanglement and vice versa.
The mathematical approach has been developed in the framework of operator mechanics, in terms of Hilbert spaces, eigenbasis sets, and tensor products.
It has been shown that the Wigner formalism provides a convenient framework that presents this theory in phase space~\cite{Ellinghaus2017}.
Here, we continue this analysis considering Coulomb entanglement from both points of view: An intuitive analogy with the language of mathematics suggests them as strong and weak formulations of the concept. 

The theoretical formulation of the Coulomb interaction and the corresponding e-e evolution equation in Wigner terms is straightforward. However, the computational aspects of the problem become prohibitive because the dimensionality doubles: The degrees of freedom of the task are defined by the coordinates of both electrons.
To address the latter, we consider two approximate models. The first one follows the ideas used in the BBGKY hierarchy of equations for many-particle systems. The approach reduces the dimensionality of the task, however, the model does not comply with the heuristic understanding of entanglement.
The formal mathematical aspects of the model are associated with a physical picture revealing what modification of the evolution rules is responsible for this.
In particular, the evolution of the individual electrons can be traced at any time. However, it is no longer coherent as the purity drops during the interaction process. This model is computationally feasible and useful if the coherence effects caused by the  Coulomb interaction are of interest. 
The second model retains the dimensionality but approximates the quantum e-e interaction with Coulomb forces. This is motivated by the fact that the latter gives rise to accelerated Newtonian trajectories, which can be computed efficiently in comparison to the complicated approach of the Coulomb-Wigner potential. Intuitively, it is expected that such a step towards classical mechanics will destroy entanglement. We prove that despite the evolution proceeding over classical trajectories, the resulting two-electron state obeys the formal definition for the entangled state. The fundamental steps of a numerical approach can be straightforwardly derived in terms of quantum particles.
Such an algorithm has 
yet to be implemented to investigate its numerical peculiarities.

In our analysis, we consider an evolution equation that is common for two alternative formulations of quantum mechanics in phase space: the standard Wigner theory and the gauge-invariant counterpart. Historically, the introduction of the Wigner function has been based on the operator's mechanical description of the fundamental problem of charged particles in electric potentials. Therefore, it was initially regarded as the auxiliary theory based on operator mechanics.  Later, with the works of Moyal and Groenewold, the Wigner theory was established as an independent, self-contained formulation of quantum mechanics \cite{Moyal49, Groenewold46}.
The formalism provides an attractive modeling approach for far-from-equilibrium mesoscale-level electron transport~\cite{Ferry2022} due to the ability
 (i) to describe quantum processes in classical terms of phase space and dynamical functions;
 (ii) to apply stochastic particle models, allowing for a natural description of the evolution of electron waves (a wave packet view provides intuitive benefits~\cite{Oriols2021});
 (iii) to treat evolution phenomena and processes of decoherence (e.g., scattering), and most importantly
 (iv) to enable a gauge-invariant generalization in terms of electromagnetic (EM) field vectors in the presence of a magnetic field.
The standard theory is based on the Weyl transform of the density matrix and the corresponding (von Neumann) evolution equation formulated in terms of the Hamiltonian and thus of the electric potential $V$. The Weyl transform involves the canonical momentum $\pv$ via a Fourier transform and defines the so-called Wigner potential  $V_{\rm w}$. The corresponding evolution equation conveniently involves the fieldless Liouville operator and thus non-accelerated Newtonian trajectories for quantum mechanical description. It is developed for the electrostatic case of vanishing magnetic fields: $\Av=0$, i.e., a zero vector potential gauge is used. 
Accordingly, it is valid in the electrostatic limit of $V$.
The Weyl formulation is apparently gauge-dependent if the magnetic field is considered. Indeed, it depends on $\Av$ via the Hamiltonian: Alternative versions of the evolution equation, depending on the choice of $\Av$ in the definition of the Wigner function, are explored by Materdey and Seyler~\cite{Materdey03n1, Materdey03n2} in the case of a homogeneous magnetic field and a symmetric gauge. In general, a gauge transform changes 
the EM potentials, while the EM field vectors, being physical quantities, remain unchanged.  
Meanwhile, however, the transform modifies the mathematical appearance of the corresponding evolution equations.
In this way, a given quantum mechanical problem is described by approaches that have different theoretical and numerical peculiarities. 
An alternative formalism - the gauge-invariant Wigner formulation of quantum mechanics - avoids this problem.  The mathematical and numerical aspects of this theory are not yet well studied \cite{Perepelkin}.
In a magnetic field, the kinetic momentum $\Pv$ of the electron is related to the canonical momentum as $\Pv=\pv-e\Av$. The former, in contrast to the latter, is a physical quantity and thus gauge-independent. In 1956, Stratonovich pursued the idea of using $\Pv$ as a variable for the phase space that defines the Wigner function $f_{\rm w}$. He derived the Weyl-Stratonovich transform, which maps the density matrix $\rho$ unitarily to $f_{\rm w}(\rv,\Pv)$ ($d$-dimensionality):
\begin{equation}
\label{WS}
    f_{\rm w}(\rv,\Pv)=\int \frac{\d\sv}{(2\pi\hbar)^d}\e^{-\frac{\rm i}{\hbar}\sv\cdot[\Pv+\frac{e}{2}\int_{-1}^1\\d\tau\Av(\rv+\frac{\sv\tau}{2})]}\rho\left(\rv+\frac{\sv}{2},\rv-\frac{\sv}{2}\right)
\end{equation}
It is interesting to note that both $\rho$ and the argument of the Fourier transform depend on the vector potential $\Av$, however, this eliminates it from the evolution equation. 
Alternative formulations of the gauge-invariant  Wigner equation for general, time-dependent inhomogeneous EM conditions were derived ~\cite{Serimaa86, Serimaa87,Levanda01}. 
The theory was recently revised to provide a pathway for practical, numerical implementations \cite{Nedjalkov2019} and further extended by a derivation of a gauge-invariant Wigner equation for linear magnetic fields ~\cite{Nedjalkov2022}. It was expressed as a Fredholm integral equation of the second kind, used to develop stochastic particle models for computing the solution~\cite{etl2024wigner, etl2024IEEE}. 
The physical settings for this case are also considered here: The two-dimensional electron evolution in the $\rv=(x,y)$-plane corresponds to transport in two-dimensional materials. The $x$-dependent magnetic field $\Bv(\rv)=(0,0, B_0+B_1x)$ with homogeneous and linear components $B_0$ and $B_1$ is normal to the plane in $z$-direction so that the Lorentz magnetic force is in the plane. The electric field $\Ev(\rv)$ has a general spatial dependence. 
A stationary electric field allows to re-express $e\Ev(\rv)$ as the electric potential energy $V$ and to reuse the Wigner potential $V_{\rm w}(\pv,\rv)$, a quantity which has been physically well analyzed over the last decades. 
We note that the near-stationary electric field conditions are not necessary but were chosen for convenience to synchronize with the standard theory. The corresponding equation is
\begin{eqnarray} \label{eq:1e}
\left(\frac{\partial}{\partial t}
+
\frac{\Pv}{m}\cdot
\frac{\partial}{\partial \rv}
+\Fv(\Pv,\rv) \cdot\frac{\partial}{\partial \Pv}  \right)f_{\rm w}\bigl(\rv, \Pv \bigr)
&=&
 \int\d\Pv' V_{\rm w}(\Pv-\Pv',\rv) 
f_{\rm w}\bigl(\rv, \Pv' \bigr)+  \\
&&\frac{B_1\hbar^2}m\frac e{12}
\left(
\frac{\partial^2}{\partial {P_y}^2}\frac{\partial}{\partial x} -
\frac{\partial}{\partial {P_x}}\frac{\partial}{\partial {P_y}}
\frac{\partial}{\partial y}\right) f_{\rm w}\bigl(\rv, \Pv), 
\nonumber
\end{eqnarray}
where $\Fv(\Pv,\rv)=\frac{e}{m}{\Pv\times\Bv(\rv)}$ is the magnetic component of the Lorentz force and
\begin{equation}
\label{WP}
{V}_{\rm w}(\rv,\Pv')=\frac{1}{{\rm i}\hbar(2\pi\hbar)^2 } \int\d\sv \e^{-\frac{\rm i}{\hbar}\sv\cdot\Pv} \left[ V\left(\rv + \frac\sv2\right) - V\left(\rv - \frac\sv2\right) \right]. 
\end{equation}
The mathematical structure of this equation unifies a variety of evolution scenarios. The term in the second row is proportional to the linear magnetic component $B_1$. The derivation of the equation, together with the assumptions and approximations, is given in \cite{Nedjalkov2022}. $B_1$  can be tuned to investigate nonlinear effects in the electron dynamics. It becomes zero if the field is homogeneous, so that the magnetic force $\Fv$ governs the characteristics of the Liouville operator on the left - the Newtonian trajectories. If the electric potential is additionally linear or quadratic, the Wigner potential \eqn{WP} gives rise to an electric force, and the evolution becomes classical. 
In the case of no magnetic field and with $V$ having a general spatial dependence, \eqn{eq:1e} becomes the evolution equation of the standard Wigner theory, while \eqn{WS} with $\Av=0$ reduces to the Weyl transform. This equation is the basis for our analysis of Coulomb entanglement.

Section~\ref{Twobody} formulates the two-electron problem and verifies the normalization of the solution and the ensuing marginal distributions. The time dependence of the process of entanglement is analyzed by using the integral form of the equation. 
In Section~\ref{Deco} we present and analyze two new approximations: reduction of the dimensionality and approximation of the Coulomb potential. The former approach utilizes the projection method used to derive the BBGKY hierarchy \cite{Balescu}. The equation formulated in the two-electron phase space gives rise to two equations that reside in the single-electron subspaces and are coupled by averaged Wigner-Coulomb potential terms.
The physical relevance of the latter compared to the genuine interaction term is discussed. It gives rise to an algorithm that resembles the widely used approach for self-consistent simulations of Boltzmann transport in semiconductor structures \cite{Vasileska2011}. 
Finally, we approximate the Coulomb potential and analyze the physical aspects of the approximation, showing that the model fulfills the orthodox definition of entanglement.
Aiming at a self-contained presentation, we include further details in the appendices.

\section{\label{Twobody} The Two-Electron Problem} 
We assume that the magnetic field generated by the charged particle's movement is negligible with respect to Coulomb repulsion.  The corresponding  Schr\"odinger equation is
\begin{eqnarray}
-{\rm i}\hbar\frac{\partial \psi(\rv_1,\rv_2)}{\partial t} = \bigg[ \frac{1}{2m}(-{\rm i}\hbar\nabla - e\Av(\rv_1))^2 + \frac{1}{2m}(-{\rm i}\hbar\nabla - q\Av(\rv _2))^2
\label{eq:2SE}
+ V(\rv_1) + V(\rv_2) + V_\textrm{int}(\rv_1, \rv_2 ) \bigg] \psi(\rv_1,\rv_2),
\end{eqnarray}
where the two electrons are indicated by the indices $1$ and $2$.
Without the Coulomb potential $V_\textrm{int}$, equation \eqn{eq:2SE} gives rise to two decoupled single-particle equations so that $\psi$ becomes a product of their solutions.

\subsection{The Two-Electron Wigner Equation}
The Wigner counterpart can be formally derived from \eqn{eq:2SE} by
following the same approach given in \cite{Nedjalkov2019}. 
The equation is written explicitly in Appendix~\ref{2EWE}.
For our analysis, it is convenient to rewrite the equation using the following short notations:  $j=1,2$ stands for the electron phase space variables if they  are arguments of operators or functions, $j=(\Pv_j,\rv_j)$, $\Fv_j$ denotes the magnetic component of the Lorentz force and ${\cal L}_j$ is the Liouville operator:
\begin{eqnarray}
&&\hskip-1cm
\label{eq:7}
f_{\rm w}(1,2)=f_{\rm w}(\rv_1, \rv_2, \Pv_1, \Pv_2);  \quad f_{\rm w1}(1)=\int\d 2f_{\rm w}(1,2),
\,\, f_{\rm w2}(2)=\int\d 1f_{\rm w}(1,2);
\\[1mm]
&&
{\cal L}_j=\frac{\Pv_j}{m}\cdot
\frac{\partial}{\partial \rv_j}
+\Fv_j
\cdot\frac{\partial}{\partial \Pv_j};\quad \Fv_j=\frac{e}{m}{\Pv_j\times\Bv(\rv_j) };\\
&&
{\cal V}_{\rm w}(j,j')=\int\d\Pv_j'\int\d\rv_j' V_{\rm w}(\Pv_j-\Pv_j',\rv_j')\delta(\rv_j-\rv_j');
 \\
&&
{\cal B}_j=\frac {B_1\hbar^2}m\frac e{12}
\left(
\frac{\partial^2}{\partial {P_y}_j^2}\frac{\partial}{\partial x_j} -
\frac{\partial}{\partial {P_x}_j}\frac{\partial}{\partial {P_y}_j}
\frac{\partial}{\partial y_j}\right); 
\end{eqnarray}
With these notations, equation \eqn{eq:Wiglinmag} can be shortly written as
\begin{eqnarray} \label{eq:Wigprod}
&&\left(\frac{\partial}{\partial t}
+
{\cal L}_1 + {\cal L}_2
\right)f_{\rm w}(1,2)
=\\[1mm]
&&{\cal V}_{\rm w}(1,1') f_{\rm w}(1',2)
+{\cal V}_{\rm w}(2,2')  f_{\rm w}(1,2')
+\bigl( {\cal B}_1+  {\cal B}_2 \bigr ) f_{\rm w}(1,2)
+ \nonumber \\[1mm]
&& \int\d\Pv_1'\d\Pv_2' {V_{\rm w}}_{\rm int}\bigl(\Pv_1-\Pv_1', \Pv_2-\Pv_2',\rv_1, \rv_2\bigr) 
f_{\rm w}(\rv_1,\rv_2, \Pv_1', \Pv_2'). 
\nonumber
\end{eqnarray}
The phase space arguments of the last term, the Coulomb-Wigner potential, which will be the focus of our analysis, are written explicitly as:
\begin{eqnarray}
\label{full}
&&{V_{\rm w}}_{\rm int}(\Pv_1,\Pv_2, \rv_1,\rv_2)=
\frac{1}{{\rm i}\hbar(2\pi\hbar)^d } \int\d\sv_1 \e^{-\frac{\rm i}{\hbar}\sv_1\cdot\Pv_1} \int\d\sv_2\e^{-\frac{\rm i}{\hbar}\sv_2\cdot\Pv_2}
\\
&&\left[ V\left(\left|\rv_1-\rv_2 + \frac{\sv_1-\sv_2}{2}\right|\right)   
- V\left(\left|\rv_1-\rv_2 - \frac{\sv_1-\sv_2}{2}\right|\right)
\right]
  .
 \nonumber
\end{eqnarray}
Here, $d$ is the dimensionality of the task. The potential is given by $V(r)= e\ln(r)/(2 \pi \epsilon_0)$ for $d=2$ and by
 $V(r) =e/(4\pi \epsilon_0r)$ for $d=3$, where $r=|\rv|$.
 
In accordance with \eqn{eq:2SE} without the Coulomb interaction, the equation decouples so that the two-electron Wigner state becomes a product of two 
solutions of  \eqn{eq:1e}.

\subsection{Normalization and Properties}
The normalization conditions are
\begin{equation}
\int\d j f_{{\rm w}j}^0(j)=1;\qquad \int\d j f_{{\rm w}j}(j)=1;\qquad \int\d 1 \d 2 f_{\rm w}(1,2)=1;\qquad j=1,2;
\label{norma}
\end{equation}
The normalization holds for any moment of the evolution time. This follows from the fact that the Wigner function, together with its derivatives, become zero at infinity.  
Details are given in Appendix~\ref{app:IF}.  In particular, the normalization of the reduced or single-electron Wigner functions $f_{{\rm w}j}$ follows from the last equality in \eqn{norma}.
The superscript $0$ in the first equality stands for the initial condition of the two-electron system $f_{\rm w}^0=f_{\rm w1}^0f_{\rm w2}^0$. We assume that the initial state corresponds to non-interacting electrons, i.e., the Coulomb interaction is \textit{switched on} at time $0$.
The physical relevance of this assumption deserves a separate analysis, however, it allows to develop a phenomenological insight into the evolution of the entanglement in accordance with \cite{EPR}.
In terms of the physical definition of the latter, the initial state is separable. We also assume that the initial electrons are sufficiently separated and are directed in a way that avoids a collision, in order to eliminate the problems with the pole of the Coulomb interaction.

The time evolution can be conveniently studied by using the integral form of \eqn{eq:Wigprod}. It links the two-electron state at a given evolution time to the initial condition via the consecutive iterations of the equation. The derivation of the integral form is discussed in Appendix~\ref{IF}.
For a small evolution interval $\Delta t$ the time integration becomes a multiplication of the integrand by $\Delta t$. The Neumann series is represented by the zeroth (the initial condition)  and the first iterative terms: 
\begin{eqnarray}\label{increment}
&&f_{\rm w}(1,2,\Delta t)= f_{\rm w1}^0(1(0))f_{\rm w2}^0(2(0))
\\&&
+\Delta t\biggl({\cal V}(1,1')f_{\rm w1}^0(1')f_{\rm w2}^0(2(0))+ {\cal V}(2,2')f_{\rm w1}^0(1(0))f_{\rm w2}^0(2')+
\biggl( ({\cal B}_1+  {\cal B}_2)f_{\rm w} \biggr ) (1(0),2(0))
\biggr)
\nonumber\\
&&
+\Delta t \int\d\Pv_1'\d\Pv_2' {V_{\rm w}}_{\rm int}\biggl(\Pv_1(0)-\Pv_1', \Pv_2(0)-\Pv_2',\rv_1(0), \rv_2(0)\biggr) 
f_{\rm w1}^0(\rv_1(0), \Pv_1',0)f_{\rm w2}^0(\rv_2(0), \Pv_2',0),
\nonumber
\end{eqnarray}
The contribution from the single-phase space operations in the second row gives rise to a separable function. This is not a consequence of the fact that it is a sum of  separable functions of the type $\phi(1)\xi(2)$. 
In general, nothing guarantees that such a sum equals a product $\Phi(1)\Xi(2)$. However, in this particular case, this can be shown with the help of the evolution equation \eqn{eq:1e}.  
The solutions of the latter, $f_{\rm w1}^{\rm n}$ and $f_{\rm w2}^{\rm n}$, for two non-interacting electrons, can be presented as corresponding Neumann expansions, as in the case of \eqn{increment}.
The expression with the first-order terms in $\Delta t$ of their product coincides with the second-row terms of \eqn{increment}. 
Thus, we identify the operator in the last row as the source of the entanglement process: First of all, the integration over the momentum coordinates mixes the variables.
Accordingly, the solution at time $\Delta t$, having the form $f_{\rm w}(\Delta t)=f_{\rm w}^{\rm n}(\Delta t)+{f_{\rm w}}_{\rm int}$, is no longer separable as the non-interacting two-electron state is corrected to the interaction counterpart $f_{\rm w}$ by  ${f_{\rm w}}_{\rm int}$. 
In the former case, we can label the electrons as first and second because it is possible to link $j$ to the corresponding initial condition $f^0_{{\rm w}j}$. 
However, this is not possible at the next evolution step when $f_{\rm w}(\Delta t)$ becomes the next initial condition: There are two sets of electron coordinates and if one of them is traced out according to \eqn{eq:7}, the resulting $f_{{\rm w}j}$ describes the other electron state, so that we can talk about one electron or the other electron, but not relate them to one of the initial states $f^0_{{\rm w}j}$.
The fact that the entanglement happens on the same time scale as the single electron evolution shows that we cannot use the time scale to discriminate the two processes by updating $f_{\rm w1}^0$ and $f_{\rm w2}^0$ independently. 


\section{ \label{Deco} Physical and Numerical Aspects}
The development of a stochastic algorithm for solving \eqn{eq:Wigprod} follows straightforward rules. The equation bears the same structure as the standard Wigner equation, which has been analyzed by using the Monte Carlo theory for solving integral equations \cite{Nedjalkov2021}.
The concepts of quantum particles that carry a sign, and are subject to generation and annihilation processes, which have already been generalized beyond the standard model, \cite{Ballicchia2024, etl2024wigner}, are independent of the fact that the number of variables doubles. Indeed, a peculiarity of Monte Carlo algorithms is their independence of the dimensionality of the task \cite{Dimov2007}. Thus, the development of a quantum particle model of the two-electron evolution governed by   \eqn{eq:Wigprod} is a feasible task and will certainly contribute to our heuristic understanding of the transport process. However, the problem is the implementation of the algorithm: A transition from a four-dimensional phase space of \eqn{eq:1e} to the eight-dimensional one of \eqn{eq:Wigprod} is an enormous challenge for available computational resources and methods. 
We first explore the idea of reducing the dimensionality of the problem via projections.

\subsection{Reduction of Dimensionality}
The decomposition of high-dimensional problems into a set of coupled low-dimensional problems is widely used in many-particle physics for the derivation of BBGKY hierarchies of equations \cite{Balescu}. 
As applied to classical electron transport, the approach allows to develop models, where the complicated distribution function of N interacting electrons is represented by N independent electrons governed by local (Newtonian) forces, updated on a short time scale by the mutual interactions. 
The e-e interaction during the time intervals between the updates is switched off. In this way, the computational effort of the problem reduces from $m^N$ to $Nm$, where $m$ is the dimensionality of the phase space of a single electron. 
This approach established itself as one of the key methods for advanced modeling of microelectronic devices and structures \cite{Vasileska2011}. For two electrons, the method is based on projections onto the single-electron subspaces and the presentation of the two-electron distribution function as a product of the single-electron counterparts.
We show that this method indeed lifts the computational challenge and can be used for studying the Coulomb force-induced correlations in the single-electron evolution and the corresponding physical quantities, but does not obey the orthodox tenet of entanglement. 
Indeed, the concept behind mixing the variables is confronted with the idea of their elimination. Nevertheless, it is insightful to consider the approach and to see how projection changes the physics of the evolution.

Equation \eqn{eq:Wigprod}, with $f_{\rm w}$ being represented by the product of the reduced Wigner functions $f_{\rm w1}f_{\rm w2}$, is integrated into the single-electron coordinates $1$ and $2$ respectively. The following set of coupled equations is obtained, where integrating over $2$ yields
\begin{subequations}\label{eq:projsys1}
\begin{align}
&\left(\frac{\partial}{\partial t}
+
{\cal L}_1 
\right)f_{\rm w1}(1)
={\cal V}_{\rm w}(1,1') f_{\rm w1}(1')
+ {\cal B}_1 f_{\rm w1}(1)
\label{eq:Wigprod_proj1a}
 +\int\d\Pv_1'
{{\cal V}_{\rm w}}_{\rm int1}(\Pv_1-\Pv_1',\rv_1)
f_{\rm w1}(\rv_1, \Pv_1'),
\\[2mm]
&{{\cal V}_{\rm w}}_{\rm int1}(\Pv_1-\Pv_1',\rv_1)=
\int\d\Pv_2\d\rv_2 \d\Pv_2' {V_{\rm w}}_{\rm int}\biggl(\Pv_1-\Pv_1', \Pv_2-\Pv_2',\rv_1, \rv_2\biggr) f_{\rm w2}(\rv_2, \Pv_2'). 
\label{eq:Wigprod_proj1b}
\end{align}
\end{subequations}
Similarly, integrating over $1$ gives
\begin{subequations} \label{eq:projsys2}
\begin{align}
&\left(\frac{\partial}{\partial t}
+
{\cal L}_2 
\right)f_{\rm w2}(2)
={\cal V}_{\rm w}(2,2') f_{\rm w2}(2')
+ {\cal B}_2 f_{\rm w2}(2) 
\label{eq:Wigprod_proj2a}
+\int\d\Pv_2'
{{\cal V}_{\rm w}}_{\rm int2}(\Pv_2-\Pv_2',\rv_2)
f_{\rm w2}(\rv_2, \Pv_2'),
\\
&{{\cal V}_{\rm w}}_{\rm int2}(\Pv_2-\Pv_2',\rv_2)=
\int\d\Pv_1\d\rv_1  \d\Pv_1' {V_{\rm w}}_{\rm int}\biggl(\Pv_1-\Pv_1', \Pv_2-\Pv_2',\rv_1, \rv_2\biggr) 
f_{\rm w1}(\rv_1, \Pv_1').
\label{eq:Wigprod_proj2b}
\end{align}
\end{subequations}
The eight-dimensional equation \eqn{eq:Wigprod} is replaced by two four-dimensional equations coupled via the reduced Wigner potential terms \eqn{eq:Wigprod_proj1b} and \eqn{eq:Wigprod_proj2b}, where the former depends on the solution of \eqn{eq:projsys2} and the latter on the solution of \eqn{eq:projsys1}.
However, the integration entirely removes the dependence of $f_{\rm w1}$ on $2$ and the dependence $f_{\rm w2}$ on $1$:  
$f_{\rm w}$ is constructed by the separable states $f_{\rm w1}(1)f_{\rm w2}(2)$.
Nevertheless, the function $f_{\rm w1}$ contains information about  $f_{\rm w2}$ or equivalently may depend on quantities defining  $f_{\rm w2}$, and vice versa. This becomes clear if one formally considers a phase space delta function, which is not an admissible quantum state but throws insight into the mathematical properties of the equations. In this way, the system accounts for correlations between $f_{{\rm w}j}$, which determine the averaged values of all physical quantities of a single electron.
To see this, we need to analyze how the projection operations modify the potential \eqn{eq:Wigprod}. 
Part of the integrals in \eqn{eq:Wigprod_proj1b} and equivalently in \eqn{eq:Wigprod_proj2b} can be performed by, e.g., using the identity 
\begin{equation}
\int_{-\infty}^\infty \d\Pv'V(\Pv-\Pv')f(\Pv')=
\int_{-\infty}^\infty \d\Pv'V(\Pv')f(\Pv-\Pv')
.
\end{equation}
The integration over $\Pv_2$ gives the electron density 
\begin{equation}
n_2(\rv,t)=\int\d\Pv f_{\rm w2}(\rv,\Pv,t).
\end{equation}
Next, we can integrate the Wigner-Coulomb potential ${V_{\rm w}}_{\rm int}$ over $\Pv_2'$. According to \eqn{full} this gives rise to $2\pi\hbar\delta(\sv_2)$, so that finally:
\begin{eqnarray}
\label{1approx}
\mathcal{{\cal V}_{\rm w}}_{\rm int1}(\Pv_1, \rv_1)=
\frac{1}{{\rm i}\hbar(2\pi\hbar) } \int\d \sv_1 \e^{-\frac{\rm i}{\hbar}\sv_1\cdot\Pv_1} 
\int\d\rv_2 n_2(\rv_2,t)\left\{ 
V\left(\left|\rv_1 + \frac{\sv_1}{2} -\rv_2\right|\right)
- V\left(\left|\rv_1 - \frac{\sv_1}{2}-\rv_2\right|\right)
\right\}
  .
\end{eqnarray}
This expression coincides with the Wigner potential of a charge distribution $n_2$. Thus, without the coupling, if $n_2$ is independent of $f_{\rm w1}$, \eqn{eq:projsys1} corresponds to the problem of a single electron evolving in an external (in general time-dependent) potential. The problem is linear, fully coherent, Markovian, and reversible. 
The Wigner and the Schr\"odinger equations provide equivalent pure state descriptions.
In particular, the corresponding wave function $\psi$ can be obtained from $f_{\rm w1}$ and vice versa  \cite{Dias2004admissible}. 
The indicator for coherence, the purity of the density matrix $\rho=\psi\psi^*$, remains unitary in time $Tr(\rho^2)=Tr(\rho)=1$.  In Appendix~\ref{Obs} we show that the coupling changes the electron dynamics entirely. The evolution, e.g., described by \eqn{eq:projsys1}, becomes both nonlinear and non-Markovian. 
The latter also characterizes the early stage of the electron decoherence process of the environment due to phonon scattering \cite{PRB2006}. The evolution of purity described by this model deserves separate analysis and will be investigated numerically elsewhere. We continue with a model that obeys the strong definition of entanglement.

\subsection{ Approximation of the Coulomb Potential}\label{app:B}
We exploit the spatial dependence of the Coulomb interaction. For convenience and without a loss of generality, all other single-particle operators, such as magnetic force and electric field, are switched off.
The Wigner potential \eqn{full}, shortly written in the two-electron phase space coordinates $\rv=\rv_1,\rv_2$ and $\Pv=\Pv_1,\Pv_2$:
\begin{eqnarray}
{V_{\rm w}}_{\rm int}(\rv,\Pv)=\frac{1}{{\rm i}\hbar(2\pi\hbar)^2 } \int\d\sv \e^{-\frac{\rm i}{\hbar}\sv\cdot\Pv} \Delta V(\rv,\sv)  \quad \textrm{with} \quad \Delta V(\rv,\sv) = V\left(\rv + \frac\sv 2\right) - V\left(\rv - \frac\sv 2\right) 
\end{eqnarray}
can be expanded into a series using
\begin{equation}
\Delta V(\rv,\sv) = V\Big(\rv + \frac{\sv}{2}\Big) - V\Big(\rv - \frac{\sv}{2}\Big) = \sum_{n=0}^\infty \frac{2}{(2n+1)!}
\left(\frac\sv 2\cdot\PD{}{\rv}\right)^{2n+1}V(\rv)
\end{equation}
%
The two-electron Wigner equation (Appendix~\ref{2EWE}) can be written as
\begin{eqnarray*} 
&&\left(\frac{\partial}{\partial t}
+
\frac{\Pv_1}{m}\cdot
\frac{\partial}{\partial \rv_1} + \frac{\Pv_2}{m}\cdot
\frac{\partial}{\partial \rv_2}\right)f_{\rm w}\bigl(\rv_1, \rv_2, \Pv_1, \Pv_2 \bigr)
=\nonumber
\\
&& \int\d\Pv_1'\d\Pv_2' {V_{\rm w}}_{\rm int}(\Pv_1-\Pv_1', \Pv_2-\Pv_2',\rv_1, \rv_2) 
f_{\rm w}(\rv_1, \rv_2,\Pv_1',\Pv_2') = \nonumber \\[4mm]
&&\frac{1}{{\rm i}\hbar(2\pi\hbar)^2 } \int \d\sv_1 \d\sv_2  \e^{-\frac{\rm i}{\hbar}\sv_1\cdot\Pv_1} \e^{-\frac{\rm i}{\hbar}\sv_2\cdot\Pv_2} \sum_{n=0}^\infty \frac{2}{(2n+1)!} \sum_{j=0}^{2n+1} {2n+1 \choose j} \\ 
 &&
 \bigg[ \bigg( \frac{\sv_1}{2}\cdot\frac{\partial}{\partial \rv_1} \bigg)^{2n+1-j}
   \bigg( \frac{\sv_2}{2}\cdot\frac{\partial}{\partial \rv_2} \bigg)^{j}
  V(|\rv_1-\rv_2|)\bigg]
\rho(\rv_1,\rv_2,\sv_1, \sv_2) 
=
\nonumber\\[4mm]
&&\frac{1}{(2\pi\hbar)^2 }\biggl (\sum_{n=0}^\infty \frac{1}{(2n+1)!}
 \bigg(\frac{i \hbar}{2}\bigg)^{2n}  \sum_{j=0}^{2n+1} {2n+1 \choose j} 
 \\
&&\bigg[ \bigg( \PD{}{\Pv_1}\cdot\frac{\partial}{\partial \rv_1} \bigg)^{2n+1-j}
   \bigg({\PD{}{\Pv_2}\cdot\frac{\partial}{\partial \rv_2} \bigg)^{j}
  V(|\rv_1-\rv_2|)\bigg]\biggr)
f_{\rm w}\bigl(\rv_1, \rv_2, \Pv_1, \Pv_2 \bigr) 
}
.\nonumber
\end{eqnarray*}
\bigskip

For the first-order approximation, when the relative coordinates are much smaller than the center of mass coordinates of the two states, holds $\Delta V(\rv_1,\rv_2,\sv_1,\sv_2)  \approx - \Fv_{{12} }\cdot \sv_1 - \Fv_{{21} }\cdot \sv_2$ with
\begin{equation}
\Fv_{{12} }(\rv_1,\rv_2)=-\frac{\partial V(|\rv_1 -\rv_2|)}{\partial\rv_1},
\qquad
\Fv_{{21} }(\rv_1,\rv_2)=-\frac{\partial V(|\rv_1 -\rv_2|)}{\partial\rv_2},
\end{equation}
so that we consider the equation
\begin{equation} 
\left[\frac{\partial}{\partial t'}
+
\frac{\Pv_1}{m}\cdot
\frac{\partial}{\partial \rv_1} 
+ \Fv_{12}(\rv_1,\rv_2)\cdot \frac{\partial}{\partial \Pv_1} 
+\frac{\Pv_2}{m}\cdot\frac{\partial}{\partial \rv_2}
+ \Fv_{21}(\rv_1,\rv_2)\cdot \frac{\partial}{\partial \Pv_2}
\right]  f_{\rm w}\bigl(\rv_1, \rv_2, \Pv_1, \Pv_2,t' \bigr) 
=0,
\label{eq:enteq}
\end{equation}
where for convenience the time argument is denoted by $t'$, and the time dependence of $f_{\rm w}$ is written explicitly. 

It is important to note that the magnetic force can be included in the force terms, and the Wigner potential operator of the external electric field can be included on the right-hand side. Our analysis is not affected by the inclusion of these single-particle operators. 
Furthermore, the Coulomb interaction gives rise to a force that depends on the two-position arguments to prevent the decomposition of the equation into a system of two decoupled equations.

In accordance with Appendix~\ref{IF}, the left-hand side  becomes a total time derivative
\eq
\label{eq:td}
\frac{\d}{\d t'} f_{\rm w}\bigl(\rv_1(t'), \rv_2(t'), \Pv_1(t'), \Pv_2(t'),t' \bigr)=0 
\eqend
over the  two-electron phase space trajectory:
\begin{eqnarray}
\label{eq:trj}
&&\rv_1(t')=\rv_1-\int^t_{t'}\frac{\Pv_1(\tau)}{m} \d\tau;\qquad
\rv_2(t')=\rv_2-\int^t_{t'}\frac{\Pv_2(\tau)}{m} \d\tau
\\
&& \Pv_1(t')=\Pv_1+\int^t_{t'}\Fv_{12}(\rv_1(\tau),\rv_2(\tau))\d\tau 
;\qquad \Pv_2(t')=\Pv_2+\int^t_{t'}\Fv_{21}(\rv_1(\tau),\rv_2(\tau))\d\tau 
\nonumber
\end{eqnarray} 
We recall that the trajectory is a line in the eight-dimensional phase space, initialized at time $t$ and parametrized by $t'$. For any $t'$, it has eight components and \eqn{eq:td} gives them grouped in the single-electron subspaces.
The integration in the interval $(0,t)$  allows to express the solution via  the initial condition:
\begin{eqnarray}
\label{eq:sol}
&&f_{\rm w}\bigl(\rv_1, \rv_2, \Pv_1, \Pv_2,t \bigr)=\\
&&f_{\rm w1}^0\left(
\rv_1-\int^t_{0}\frac{\Pv_1(\tau)}{m} \d\tau,\,\,
\Pv_1-\int^t_{0}\Fv_{12}(\tau)\d\tau \right)
f_{\rm w2}^0\left(\rv_2-\int^t_{0}\frac{\Pv_2(\tau)}{m} \d\tau,\,\,
\Pv_2-\int^t_{0}\Fv_{21}(\tau))\d\tau \right)
\nonumber
\end{eqnarray}
Equation \eqn{eq:sol} shows that at any time the solution $f_{\rm w}$ is again a product of two functions. Besides, the way the equation is written suggests that the latter are functions of the corresponding single-electron coordinates, and hence the two-electron state is separable at any time. Fortunately, \eqn{eq:trj} shows that any of these functions depend on the full set of coordinates of the two-electron phase space. 
This can already be seen at a small time step after the interaction is switched on:
\begin{eqnarray}
\label{eq:solinc}
&&f_{\rm w}\bigl(\rv_1, \rv_2, \Pv_1, \Pv_2,\Delta t \bigr)=\\
&&f_{\rm w1}^0\left(
\rv_1-\Delta t\frac{\Pv_1}{m}, \,\,
\Pv_1-\Delta t \Fv_{12}(\rv_1,\rv_2)\right)
f_{\rm w2}^0\left(\rv_2-\Delta t\frac{\Pv_2}{m},\,\,
\Pv_2-\Delta t \Fv_{21}(\rv_1,\rv_2) \right)
\nonumber
\end{eqnarray}
We come to the important conclusion that the entanglement happens due to the dependence of the force on both spatial coordinates. If it was a constant or had the particular dependence $\Fv_{12}(\rv_1)$ and $\Fv_{21}(\rv_2)$, then $f_{\rm w}$ evolves as a separable state.

Remarkably, classical trajectories provide a heuristic insight into such quantum processes as entanglement. It is a feature of the Wigner formalism in cases when evolution rules are classical, but the initial state is quantum. Further understanding comes from the fact that integrating $f_{\rm w}$ on a single-electron phase space gives the reduced states $f_{\rm w1}(1)$ or $f_{\rm w2}(2)$ that identify two electrons at time $t$.
However, neither of them can be associated solely to $f_{\rm w1}^0$ or  $f_{\rm w2}^0$:
The reduced states depend both on $f_{\rm w1}^0$ and  $f_{\rm w2}^0$.

The numerical approach to \eqn{eq:sol}  essentially relies on the construction of the Newtonian trajectories: There is a lot of experience with methods for particle transport modeling \cite{Vasileska2011}. The problem related to the high dimensionality lends itself to parallel solution algorithms, which are widely available, and will be further explored in future work.  

\section{Discussion}


The Wigner equation for two electrons coupled by Coulomb interaction evolving in a plane already depends on eight phase space coordinates,
as required, e.g., in the presence of magnetic fields.  
The problem poses enormous computational challenges and requires approximate models. We present and analyze two such models based on (i) the BBGKY reduction of variables and (ii) the force approximation of the Coulomb interaction. 
The former gives rise to a separable state, a product of two functions of the single-electron coordinates. However, the coupled single-electron dynamics are non-linear and non-Markovian and thus very different from a pure state evolution. 
The second model obeys the strong definition of entanglement. The two-electron state is still a product of two functions. 
However, the separability of the initial state is already lost at the very beginning of the interaction and is linked to the fact that the Newtonian trajectories belong to the two-electron phase space.
The integral representation of the evolution equations throws intuitive and physical insight into the process of entanglement and the peculiarities of the approximate models.

The considered models pursue the computational ability to model the evolution of two interacting electrons and the process of their entanglement.  The development of efficient numerical approaches is only one aspect of future work. First, we need to analyze whether these approximate models are relevant to provide  the physics, which is in the focus of the particular view to the process.
The loss of coherence of the reduced states is a basic indicator for deviating from a single-particle pure state evolution. 
Thus, the time evolution of the purity computed with the two models and compared with the full two-electron problem will shed insight into the problem. 

\section{Acknowledgment}
This research was funded in whole by the Austrian Science Fund (FWF), contract   P37080-N.



\clearpage

\clearpage

\appendix

\subsection{\label{2EWE} The Two-Electron Wigner Equation for a Linear Magnetic Field.}

Straightforward calculations, which merely follow \cite{Nedjalkov2022}, lead to a long, but human-friendly equation:
\begin{eqnarray} \label{eq:Wiglinmag}
&&\hskip-10mm
\left(\frac{\partial}{\partial t}
+
\frac{\Pv_1}{m}\cdot
\frac{\partial}{\partial \rv_1} + \frac{\Pv_2}{m}\cdot
\frac{\partial}{\partial \rv_2}
+\frac{e}{m}{\Pv_1\times\Bv(\rv_1)} \cdot\frac{\partial}{\partial \Pv_1} + \frac{e}{m}{\Pv_2\times\Bv(\rv_2)}
\cdot\frac{\partial}{\partial \Pv_2}\right)f_{\rm w}\bigl(\rv_1, \rv_2, \Pv_1, \Pv_2 \bigr)
=\nonumber
\\
&& \int\d\Pv_1' V_{\rm w}(\Pv_1-\Pv_1',\rv_1) 
f_{\rm w}(\rv_1, \rv_2,\Pv_1',\Pv_2) + \int\d\Pv_2' V_{\rm w}(\Pv_2-\Pv_2',\rv_2) 
f_{\rm w}(\rv_1, \rv_2,\Pv_1,\Pv_2') +
\label{lform} \nonumber \\ 
&& \int\d\Pv_1'\d\Pv_2' {V_{\rm w}}_{\rm int}(\rv_1, \rv_2,\Pv_1-\Pv_1', \Pv_2-\Pv_2') 
f_{\rm w}(\rv_1, \rv_2,\Pv_1',\Pv_2') + \\
&&\frac {B_1\hbar^2}m\frac e{12}
\left(
\frac{\partial^2}{\partial {P_y}_1^2}\frac{\partial}{\partial x_1} -
\frac{\partial}{\partial {P_x}_1}\frac{\partial}{\partial {P_y}_1}
\frac{\partial}{\partial y_1}\right) f_{\rm w}\bigl(\rv_1, \rv_2, \Pv_1, \Pv_2 )  +  \nonumber \\
&&\frac {B_1\hbar^2}m\frac e{12}
\left(
\frac{\partial^2}{\partial {P_y}_2^2}\frac{\partial}{\partial x_2} -
\frac{\partial}{\partial {P_x}_2}\frac{\partial}{\partial {P_y}_2}
\frac{\partial}{\partial y_2}\right) 
f_{\rm w}\bigl(\rv_1, \rv_2, \Pv_1, \Pv_2 ). 
\nonumber
\end{eqnarray}

For non-interacting distinguishable electrons, the two-electron Wigner function is a product of the two single-electron Wigner states: $f_{\rm w}\bigl(\rv_1, \rv_2, \Pv_1, \Pv_2) =f_{\rm w1}\bigl(\rv_1,  \Pv_1) f_{\rm w2}\bigl(\rv_2,  \Pv_2)$, solutions of \eqn{eq:1e}.


\subsection{ Integral Features}\label{app:IF}
A feature of the operators involved in the single-particle equation {\eqn{eq:1e}} is that their integration over the phase space variables $\textstyle\int\d\Pv \d\rv$ gives zero. 
The assumption is that the Wigner function, together with its derivatives, become zero at infinity. 
For the Wigner potential operator, this property is directly related to the definition of $V_{\rm w}$, which is a Fourier transform of the difference $V(\rv+\sv)-V(\rv-\sv)$.
For the operators involving derivatives in one or more components of these variables holds:
\begin{eqnarray*}
    \int\d\Pv_2 \d\rv_2 \frac{\Pv_2}{m} \cdot \frac{\partial}{\partial \rv_2}  f_{\rm w}\bigl(\rv_1, \rv_2, \Pv_1, \Pv_2 ) = \int\d\Pv_2  \frac{\Pv_2}{m} f_{\rm w}\bigl(\rv_1, \rv_2, \Pv_1, \Pv_2 ) \bigg|_{\rv_2=-\infty}^{\infty} = 0, 
\end{eqnarray*}
where we used that $f_{\rm w}(\rv_1, \pm\infty, \Pv_1, \Pv_2 )=0 $.
In the same way, we can demonstrate that
\begin{eqnarray*}
&&
\int \d\rv_2 \d\Pv_2 \d\Pv_2' V_{\rm w}(\Pv_2-\Pv_2',\rv_2) f_{\rm w}(\rv_1, \rv_2,\Pv_1,\Pv_2') = \\
&&
\int \d\rv_2 \d\Pv_2' \left( \int V_{\rm w}(\Pv_2-\Pv_2',\rv_2) \d\Pv_2 \right) f_{\rm w}(\rv_1, \rv_2,\Pv_1,\Pv_2') = 0,
\end{eqnarray*}
because the internal integral on $\Pv_2$ is zero.
We can also demonstrate that 
\begin{eqnarray*}
&&
\int\d\rv_2 \d\Pv_2 \frac{e}{m}{\Pv_2\times\Bv(\rv_2)}
\cdot\frac{\partial}{\partial \Pv_2}f_{\rm w}\bigl(\rv_1, \rv_2, \Pv_1, \Pv_2 \bigr) = \\
&&
\int\d\rv_2 \d\Pv_2 \Bigg(\frac{e}{m}{{P_y}_2 B(\rv_2)}
\cdot\frac{\partial}{\partial {P_x}_2} - \frac{e}{m}{{P_x}_2 B(\rv_2)}
\cdot\frac{\partial}{\partial {P_y}_2}\Bigg)f_{\rm w}\bigl(\rv_1, \rv_2, \Pv_1, \Pv_2 \bigr) = 0,
\end{eqnarray*}
using that $f_{\rm w}\bigl(\rv_1, \rv_2 , \Pv_1, \pm \infty )=0$.\\


\subsection{\label{IF} Integral Form}

The characteristics of the Liouville operator  ${\cal L}_j=\frac{\partial}{\partial t'} +{\cal L}(\rv_j,\Pv_j)$ (see \eqn{eq:Wigprod}) are the Newtonian trajectories 
\eq
\rv_j(t')=\rv_j-\int^t_{t'}\frac{\Pv_j(\tau)}{m} \d\tau;\qquad \Pv_j(t')=\Pv_j+\int^t_{t'}\Fv_j(\tau)\d\tau 
\label{traj}
\eqend
initialized at time $t$ by $\rv_j,\Pv_j$ and shortly denoted by $j(t')$. The parametrization time $t'$ runs backward, in the sense that $t>t'$.
This can be used to express the left-hand side of the equation as a total time derivative \cite{Nedjalkov2021}:
\begin{eqnarray} \label{eq:Wigtwo}
&&\frac{\d}{\d t'}f_{\rm w}(1(t'),2(t'))
=\int\d 1'{\cal V}_{\rm w}(1(t'),1') f_{\rm w}(1',2(t'))\\[2mm]
&&
+\int\d 2'{\cal V}_{\rm w}(2(t'),2')  f_{\rm w}(1(t'),2')
+\biggl( ({\cal B}_1+  {\cal B}_2)f_{\rm w} \biggr ) (1(t'),2(t'))
 \nonumber \\[2mm]
&& + \int\d\Pv_1'\int\d\Pv_2' {V_{\rm w}}_{\rm int}\biggl(\Pv_1(t')-\Pv_1', \Pv_2(t')-\Pv_2',\rv_1(t'), \rv_2(t')\biggr) 
f_{\rm w1}(\rv_1(t'),\rv_2(t'), \Pv_1' ,\Pv_2'). 
\nonumber
\end{eqnarray}
After an integration over $t'$ in the interval $(0,t)$ we obtain a Fredholm integral equation of the second kind with the initial condition $f_{\rm w}^0=f_{\rm w1}^0(1(0))f_{\rm w2}^0(2(0))$ as the free term.
In particular, the trajectory initialized by $j(t'=t)$ defines the phase space point $j(t'=0)$ of the initial condition. For convenience, we explicitly write the integral form of the single-electron counterpart, obtained in the same way from \eqn{eq:1e}:

\begin{eqnarray} 
\label{eq:Wig1int}
f_{\rm w}(\rv,\Pv,t)
=&&f_{\rm w}^0(\rv(0),\Pv(0)) 
\\
&&+
\int\limits_0^t\d t' \left\{\int\d\Rv' V_{\rm w}(\Pv(t')-\Pv',\rv(t')) f_{\rm w}(\rv(t'),\Pv',t')
+\bigl({\cal B}f_{\rm w} \bigr ) (\rv(t'),\Pv(t'),t')\right\}
\nonumber
\end{eqnarray}
The two-electron equation has the same structure, except that the Coulomb interaction term from \eqn{eq:Wigtwo} appears in the curly brackets for the time integral.
A consecutive replacement of a Fredholm integral equation into itself expresses the solution as a series of consecutive integrals of the equation's kernel on the initial condition, also called the Neumann expansion of the solution.

\subsection{\label{Obs} Analysis of the Reduced Model}

The integral form of system \eqn{eq:projsys1} and \eqn{eq:projsys2} is obtained by following the approach of Appendix~\ref{IF}. We can skip all single-electron operators and consider only the e-e interaction potential:
\begin{eqnarray}
\label{coup1}
f_{\rm w1}(\rv_1,\Pv_1,t)
=&&f_{\rm w1}^0(\rv_1(0),\Pv_1(0))
+
\int\limits_0^t\d t' \int\d\Pv_2 \d\rv_2 \d\Pv_1' \d\Pv_2'
\\
&&
{V_{\rm w}}_{\rm int}\biggl(\Pv_1(t')-\Pv_1', \Pv_2-\Pv_2',\rv_1(t'), \rv_2\biggr)
f_{\rm w2}(\rv_2, \Pv_2',t')
f_{\rm w1}(\rv_1(t'), \Pv_1',t')
\nonumber
\end{eqnarray}
\begin{eqnarray}
\label{coup2}
f_{\rm w2}(\rv_2,\Pv_2',t')
=&&f_{\rm w2}^0(\rv_2(0),\Pv_2'(0))
+
\int\limits_0^{t'}\d t'' \int  \d\Pv \d\rv \d\Pv_1'' \d\Pv_2''
\\&&
{V_{\rm w}}_{\rm int}\biggl(\Pv-\Pv_1'', \Pv_2'(t'')-\Pv_2'',\rv, \rv_2(t'')\biggr)
f_{\rm w2}(\rv_2(t''), \Pv_2'',t'')
f_{\rm w1}(\rv, \Pv_1'',t'').
\nonumber
\end{eqnarray}
Here the trajectories $1(t')$ and $2'(t'')$ are initialized in \eqn{traj}  by $\rv_1,\Pv_1,t$ and $\rv_2,\Pv_2',t'$ respectively. 
In the second equation, the arguments of $f_{\rm w2}$ are chosen in accordance with the way they appear in the first equation, and the phase space integration variables are renamed for convenience. 
The solution of equation \eqn{coup1} can be presented as a series in two ways: the regular Neumann expansion, or with the help of the Neumann expansion of $f_{\rm w2}$ as a result of the coupling between the two equations.
A direct replacement of \eqn{coup2} into \eqn{coup1} allows us to rewrite the latter as a sum of several terms: 

\begin{eqnarray}
\label{coupterm}
&&\int_0^t\d t' \int_0^{t'}\d t''
\int  \d\Pv \d\rv \d\Pv_2 \d\rv_2 \d\Pv_1' \d\Pv_2'\d\Pv_1'' \d\Pv_2''
\\
&& 
{V_{\rm w}}_{\rm int}\biggl(\Pv_1(t')-\Pv_1', \Pv_2-\Pv_2',\rv_1(t'), \rv_2\biggr)
{V_{\rm w}}_{\rm int}\biggl(\Pv-\Pv_1'', \Pv_2(t'')-\Pv_2'',\rv, \rv_2(t'')\biggr)
\nonumber\\
&& 
f_{\rm w2}(\rv_2(t''), \Pv_2'',t'')
f_{\rm w1}(\rv, \Pv_1'',t'')
f_{\rm w1}(\rv_1(t'), \Pv_1',t')
\nonumber
\end{eqnarray}
It is important to note that this is not a term of the Neumann expansion for $f_{\rm w1}$, but a part of the integral kernel defining the equation. Indeed, it is not obtained by iterative replacement of \eqn{coup1} into itself, but with the help of \eqn{coup2}. 
Without a coupling, namely if \eqn{coupterm} is zero, then $f_{\rm w2}$ evolves independently of $f_{\rm w1}$, and equation \eqn{coup1} becomes a Wigner equation with a potential \eqn{full}, which determines a Schr\"odinger type (linear, coherent, Markovian) evolution.
The term \eqn{coupterm} entirely changes the dynamics of the state $f_{\rm w1}$. It becomes nonlinear and non-Markovian, i.e., an initial condition is not sufficient to determine the evolution, but it depends on the whole history of the state, as indicated by the nested time integrals.
This is the situation with the Levinson and Barker-Ferry equations, which describe ultrafast electron-phonon dynamics and recover Boltzmann scattering in the long time limit \cite{PRB2006}.

\clearpage

\bibliographystyle{unsrt}
\bibliography{references}

\end{document}